\documentclass[12pt]{article}
\usepackage[UKenglish]{babel}
\usepackage{algorithm} 
\usepackage{algpseudocode}
\usepackage{amsmath,amssymb,bm,mathrsfs}
\usepackage{graphicx}
\usepackage{color}
\usepackage{subfig}
\usepackage[numbers]{natbib}
\usepackage{authblk}
\bibpunct{(}{)}{;}{a}{}{,}

\renewcommand{\Re}{\mathbb{R}}
\begin{document}
\title{Enhanced Archaeological Predictive Modelling \\ in Space Archaeology}
\author[1]{Li Chen }
\author[1]{Carey E. Priebe }
\author[1]{Daniel L. Sussman}
\author[2]{Douglas C. Comer }

\author[3]{Will P. Megarry}
\author[4]{James C. Tilton}
\affil[1]{Department of Applied Mathematics and Statistics, Johns Hopkins University, Baltimore, MD 21218, USA }
\affil[2]{ Cultural Site and Research Management, Baltimore, MD 21218, USA}
\affil[3]{School of Archaeology, University College Dublin, Belfield, Dublin 4, Republic of Ireland}
\affil[4]{NASA Goddard Space Flight Center, Mail Code 606.3, Greenbelt, MD, 20771, USA}

\date{\today}
\maketitle

\begin{abstract}
Identifying and preserving archaeological sites before they are destroyed is a very important issue. In this paper, we develop a greatly improved archaeological predictive model $APM_{enhanced}$ that predicts where archaeological sites will be found. This approach is applied to remotely-sensed multispectral bands and a single topographical band obtained from advanced remote sensing technologies such as satellites and Airborne Laser Scanning (ALS). Our $APM_{enhanced}$ is composed of band transformation, image analysis, feature extraction and classification. We evaluate our methodology on the sensor bands over Ft.\ Irwin, CA, USA. A nested bi-loop cross-validation and receiver operating characteristics curves are used to assess the performance of our algorithm. We first validate our method on the east swath of Ft.\ Irwin and then test on a separate dataset from the west swath of Ft.\ Irwin. A convex combination of two methodologies: $APM_{conventional}$, which has been used among archaeologists for many years, and our $APM_{enhanced}$, is demonstrated to yield superior classification performance compared to either alone at low false negative rates. We compare the performance of our methodology on different band combinations, chosen based on the archaeological importance for these sensor bands. We also compare the two types of $APM$s in the aspects of input data, output values, practicality and transferability.

\end{abstract}

\section{Introduction}
Archaeological sites are uniquely important sources of knowledge relevant to history, culture and nature. They convey messages from the past and embed the link between people of this modern world with people who inhabited the same areas hundreds or thousands of years ago. From a pure scientific standpoint, these sites are extremely valuable for studying the patterns of biological variations among humans and their ancestors, human achievements and linguistic origins.  Protecting archaeological sites means guarding the archaeological heritage to which human beings are entitled. However, archaeological sites are being destroyed at an incredible rate. They are faced with many threats such as road constructions, developments in urban and rural areas, mining, and agriculture \citep{destruction}. Growing populations are crowding into areas not previously occupied by human beings. Once destroyed, an archaeological site cannot be excavated. It is lost forever because archaeological research depends upon finding uncontaminated material in original context. Sites are often destroyed before humankind becomes aware of their presence. Identification and preservation of archaeological sites against damage and destruction is remarkably important. 

Traditional archaeological predictive models ($APM_{conventional}$) are intended to identify regions within which sites are likely to be found and deal only with how suitable a region is for a certain activity rather than actually finding sites. Development of $APM_{conventional}$ began approximately 50 years ago \citep{willey1953prehistoric}. It is an  ``inductive'' method derived from observations rather than from theory; for instance the environmental factors in $APM_{conventional}$ are selected by archaeologists based on experience and expertise. It is a region-based predictive approach that incorporates factors including slope, vegetation cover, proximity to water, elevation and so on. Some $APM_{conventional}$ models may be formed from a successful but informal hypothesis and later examined under simple statistical procedures. Often, archaeologists delineate regions by testing for frequency of sites against a random distribution. 

\begin{figure} 
  \centering
  \includegraphics[width=4 in]{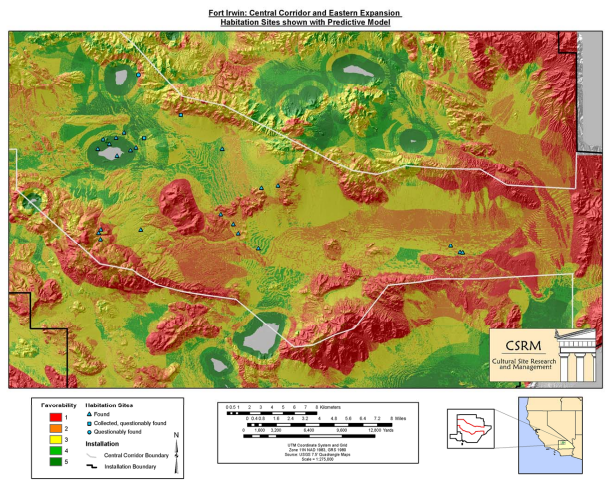}
  \caption[An example of $APM_{conventional}$ for Ft.\ Irwin sites. The traditional APM has identified regions (areas) where archaeological sites are more or less likely to be found.  
In the APMs for Fort Irwin, dark green areas are regions where sites aren't likely to be found, red areas where sites are least likely, and colors in between ranked accordingly.]
   {An example of $APM_{conventional}$ for Ft.\ Irwin sites. The traditional APM has identified regions with which archaeological sites are more or less likely to be found.  
In the APMs for Fort Irwin, dark green areas are regions where sites are likely to be found, red areas where sites are least likely, and colors in between ranked accordingly.}
   \label{fig:apm_conv} 
\end{figure}

A mathematical formulation of a traditional APM is given by a function

\begin{equation}
APM_{conventional}: \mathbb{R}^2 \to \mathbb{R}
\end{equation}
\begin{equation}
APM_{conventional}: s \mapsto t. 
\end{equation}
The input $ s = (x, y) \in \{1,2,...,w\}\times\{1,2,...,h\}  \in \mathbb{R}^2$ is the geographic location of a site on an $\mathbb{R}^2$ map of pixels, where we can consider $x$ as the row number and $y$ as the column number in the map. The output $t$ gives discretized and usually integer scores in $\mathbb{R}$ for regions, where higher scores suggest high possibility of archaeological regions and lower scores suggest regions where sites are less likely to be found. Fig [\ref{fig:apm_conv}] is an example of $APM_{conventional}$ for archaeological habitation sites in Ft.\ Irwin.  Regions where sites are most likely to be found are dark green, regions where sites are least likely to be found are red, and regions of intermediate likelihood are color coded accordingly.

Because geological characteristics vary among regions, $APM_{conventional}$ for different maps can have different influencing factors. Transferability of $APM_{conventional}$ is limited in the sense that archaeologists would always need to survey the field in a new region that depends on different archaeological factors. This is also very costly and time-consuming. Advanced technologies such as airborne and spaceborne satellites make possible the use of sensor bands to directly detect archaeological sites \citep{key-1}. Analysis of these sensor bands is one effective solution for site identification because localized environmental changes from long ago have persisted to the present due to frequent human activities in the past. We intend to construct an enhanced $APM_{enhanced}$ using sensor bands \citep{chen,key-3} with high transferability and accuracy. Such APMs can tremendously benefit the progress of preserving archaeological sites and save the cost and time of on-site investigation. Moreover, it is desirable to build enhanced APMs that yield low false negative rates. Failing to detect sites before construction projects begin, for example, can produce cost overruns when projects are delayed, and might even necessitate abandoning the project.

In this article, we develop an improved archaeological predictive model
\begin{equation}
APM_{enhanced}: \mathbb{R}^2 \to \mathbb{R}
\end{equation}
\begin{equation}
APM_{enhanced}: s \mapsto t 
\end{equation}
to classify archaeological sites where $ 1 $ denotes a site of archaeological significance and $ 0 $ denotes otherwise. 

The original dataset has two swaths (east and west) of Ft.\ Irwin military reservations, CA, USA. The dataset for each swath consists of eight multispectral bands from the WorldView-2 satellite and one slope band from Airborne Laser Scanning (ALS) \citep{2000light} data. Based on the original nine bands, we construct four additional tassel cap (KTT) bands \citep{yarbrough2005using}
and one normalized difference vegetation index (NDVI) band. 
We intend to classify archaeological sites from non-sites. 

We first train our method on 36 band difference ratios obtained from the eight multispetral bands and one slope band from east Ft.\ Irwin. To validate $APM_{enhanced}$ on the training set, we use a nested leave-one-out cross validation where the inner loop identifies the best PCA dimension, and the outer loop uses the selected PCA dimension to classify the left-out sample. We assess the performance on the eastern portion of Ft.\ Irwin via the receiver operating characteristic (ROC) curve. We particularly use the area under the curve (AUC) as a summary statistic to evaluate the performance of our approach \citep{bradley1997use}\footnote{We are aware of the controversies of using this statistic and use this evaluation with caution.}. Our methodology on the training set achieves both high accuracy and high true negative rate for low false negative rate for archaeological site classifications. We then test our method on the west swath of Ft.\ Irwin. We compare  $APM_{enhanced}$  with  $APM_{conventional}$  and find that  $APM_{enhanced}$  outperforms the  $APM_{conventional}$  in both training and testing sets, especially for low false negative rates. We also consider a convex combination model:

\begin{equation}
 APM_{\gamma} = (1-\gamma) APM_{conventional} + \gamma APM_{enhanced}
\end{equation}
and show that for the eastern region, $APM_{\gamma}$ outperforms both $APM_{conventional}$ and $APM_{enhanced}$. This is especially significant because the $APM_{conventional}$ has been evaluated as being high-performing as measured by the standard metric generated for this purpose by archaeologists, the ``gain statistic'' \citep{kvamme1}. Then we use our method as a protocol for model selection by comparing the classification accuracies of APMs using 36, 45, 66, and 78 band difference ratios. The comparison analysis of band combinations not only demonstrates the bands' levels of archaeological relevance to classification, but also demonstrates that a wise choice of features can help improve prediction accuracy.  

The remainder of the article is organized as follows. In Section 2, we describe the data set and present $APM_{enhanced}$ algorithm in detail. Section 3 shows the results of training and testing on east and west swaths of Ft.\ Irwin, and compares the performance of $APM_{conventional}$, $APM_{enhanced}$ and the convex model $APM_{\gamma}$. In Section 4, we apply $APM_{enhanced}$ to different dimensions of band difference ratios and summarize a comparison between $APM_{conventional}$ and $APM_{enhanced}$. These modelling approaches have great applicability to both the management and preservation of archaeological sites and archaeological research \citep{doug1, doug2, menze}. In this article, we demonstrate that a clever architecture, which combines machine learning technique and human expertise, yields improved performance.
 
\section{The WorldView-2 and the ALS Data}

\begin{figure}
  \centering
  \includegraphics[width=4in]{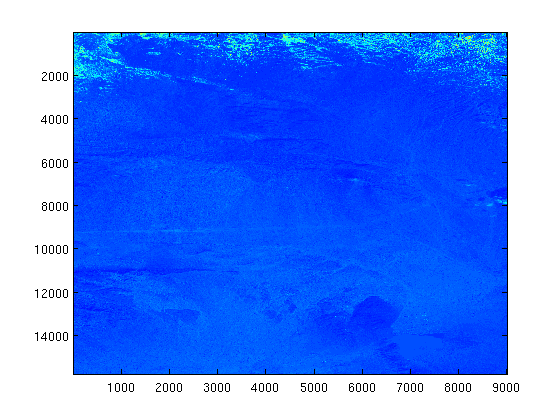}
  \caption[Image of Coastal Blue Sensor Band]
   {Image of Coastal Blue Sensor Band}
   \label{fig:sample_band}
\end{figure}

\begin{figure} 
  \centering
  \includegraphics[width=4 in]{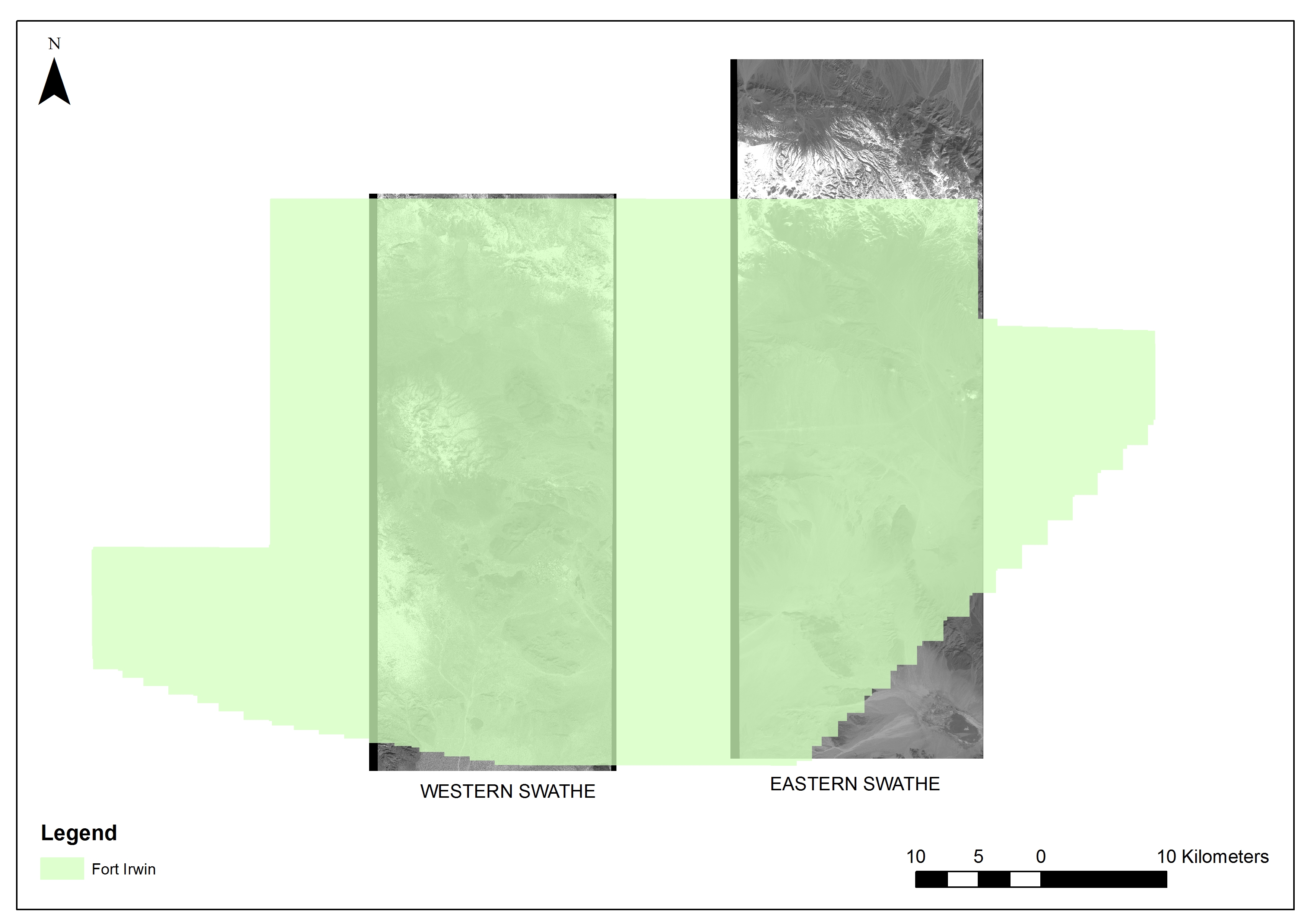}
  \caption[Eastern and western swathes of Ft.\ Irwin]
   {The east and west swathes of Ft.\ Irwin.}
   \label{fig:wavelength} 
\end{figure}

The data in our analysis has nine remotely sensed bands \citep{Aerial}. Eight of them are obtained from the WorldView-2 satellite\footnote{WorldView-2 is the second next-generation high-resolution satellite of DigitalGlobe, Inc., Longmont, CO, USA (http://www.digitalglobe.com). The satellite has eight spectral sensors in the near infra-red range.} imagery with bands: coastal blue, blue, green, yellow, red, red edge, near infrared I, near infrared II. An image of the coastal blue band is seen in Fig [\ref{fig:sample_band}]. The ninth band: slope, is calculated from the Airborne Laser Scanning (ALS) data collected by Department of Defense sensors. Our study analyzes two swaths of territory\footnote{There are fourteen swaths of WorldView-2 satellite imagery data covering the entire land area of China Lake and Ft.\ Irwin military reservations in California.}, where the first swath covering a large part of the eastern portion of Ft.\ Irwin, collected on Sep \ 22, \ 2011, is used as a training set and the second swath covering the western portion of Ft.\ Irwin, collected on Dec \ 30, \ 2010, is used as a testing set. These 8 spectral bands were orthorectified to 2-meter ground resolution with 11- bit radiometric resolution (stored as 16-bit integers). We form an eight spectral band data set with 9859 columns and 61098 rows\footnote{DigitalGlobe provided the swath of the western Ft.\ Irwin data in 45 separate 4096 $\times$ 4096 pixel blocks of data originally.}.  
All locations outside of the Ft.\ Irwin land area boundary were masked out, as shown in Fig [\ref{fig:wavelength}].

\begin{table}
\caption{WorldView-2 spectral bands.}
\begin{tabular} {| l |  p{5cm}  |  p{4cm}  | l |} \hline
\textbf{Band} & \textbf{Lower and Upper Band Edge (nm)} & \textbf{Center Wavelength (nm)} \\\hline
\textbf{Coastal Blue} & 396 and 458 & 427 \\\hline
\textbf{Blue }& 442 and 515 & 478 \\\hline
\textbf{Green} & 506 and 586 & 546 \\\hline
\textbf{Yellow} & 584 and 632 & 608 \\\hline
\textbf{Red} & 624 and 694 & 659 \\\hline
\textbf{Red Edge} & 699 and 749 & 724 \\\hline
\textbf{Near Infrared 1} & 765 and 901 & 833 \\\hline
\textbf{Near Infrared 2} & 856 and 1043 & 949 \\\hline
\end{tabular}
\end{table}

The slope data used in this study is derived from ALS data \citep{2000light} collected by the Department of Defense. It consists of images created from the data orthorectified to 2-meter ground resolution. The data ranges in integer values from 0 to 86 degrees, where 90 degrees indicates a sheer cliff. The slopes for the east Ft.\ Irwin and west Ft.\ Irwin have different sizes from the WorldView-2 multispectral bands. Thus we crop the multispectral bands data to 
a 9859-column-by-23000-row subset data that covered the grounds of a western portion of Ft.\ Irwin for the testing set, and a 9002-column-by-15780-row subset of the eastern swath was used as the training set so that all the sensor bands have a consistent size. Although the slope band and multispectral bands are generated from different devices, we made sure they are registered for the correct site locations.

We also consider four additional tassel cap bands (KTT): brightness, greenness, wetness and the fourth KTT band \citep{yarbrough2005using}, all of which were linear combinations of the spectral bands. In Section 4,  we discuss four types of band combinations that include the KTT and NDVI bands, and compare the performances of $APM_{enhanced}$ on these combinations. 

\begin{figure}
  \centering
  \includegraphics[width=6in]{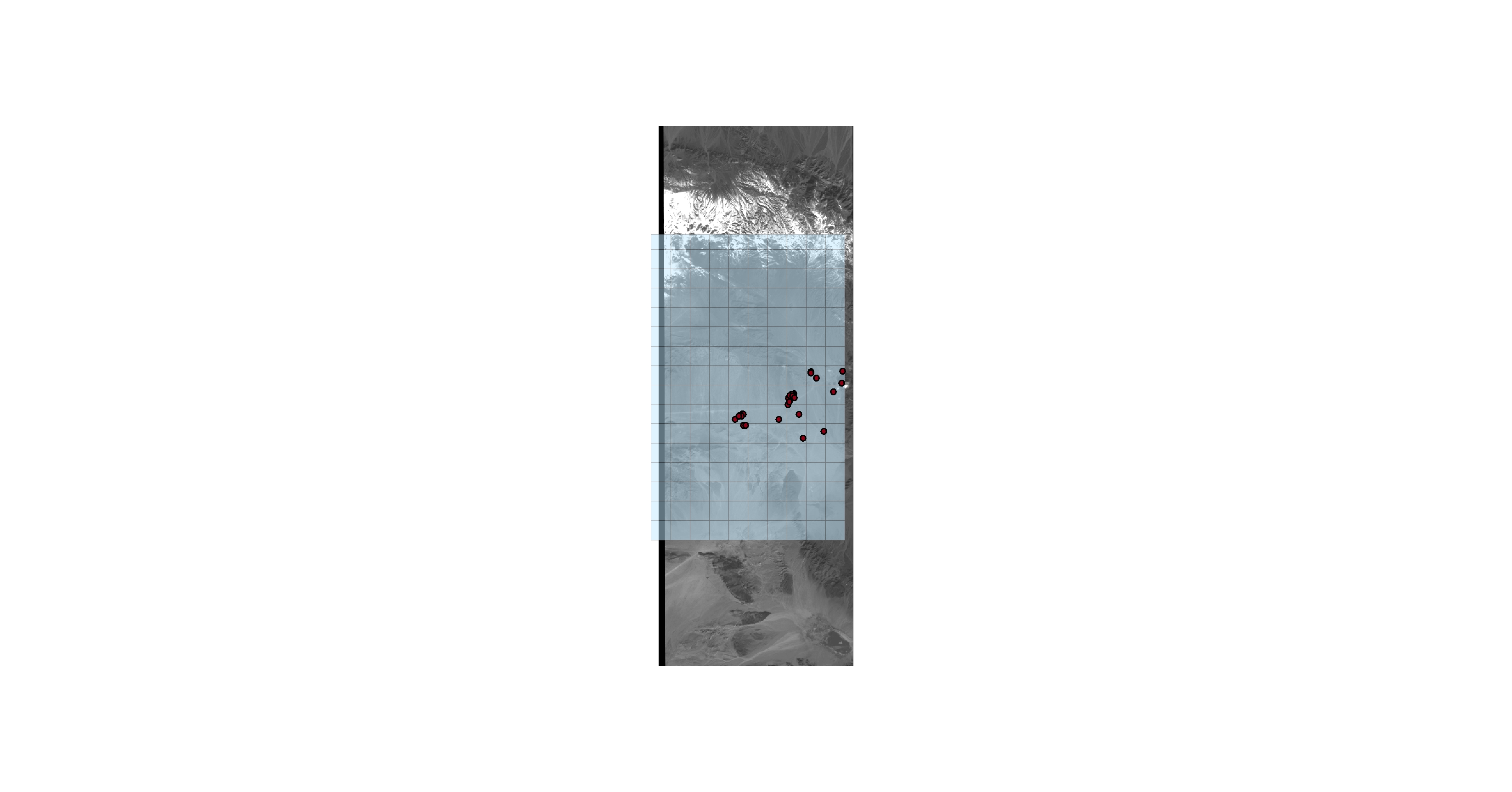}
  \caption[WorldView2 Eastern Ft.\ Irwin Data with found sites]
   {Site locations in eastern Ft.\ Irwin swath. The light blue gridded rectangular patch is the subset of the original data used in our analysis. The data is 2-meter resolution, which means there is a point every 2 metres. For Ft.\ Irwin swaths, there are too many points to process into surface models like elevation and slope so the data is distributed in grids as shown in this figure.}
\end{figure} 

\section{The Enhanced Archaeological Predictive Model} \label{sec: apm_enhanced}
Our method $ APM _{enhanced} $ is a composition of four functions $ h \circ g \circ f \circ \mathcal{D}$, where 
\begin{equation}
\mathcal{D} : \mathbb{R}^2 \to \mathbb{R}^2, \text{ as band transformation}
\end{equation}
\begin{equation}
f:\mathbb{R}^2 \to \Re^{\tilde{d}}, \text{ as image processing} \label{eq:imgProcessor3}
\end{equation}
\begin{equation}
g:\mathbb{R}^{\tilde{d}}\to\mathbb{R}^{d} \text{ with } d<\tilde{d}, \text{ as feature extraction} \label{eq:featExtr} 
\end{equation}
\begin{equation}
h:\mathbb{R}^{d}\to [0,1], \text{ as classification} \label{eq:classifier}
\end{equation}

\subsection{The Band Difference Ratio Applied to Data Bands} 

Each site $ s = (x, y) $ is a point in $\mathbb{R}^2$, where we can consider $x$ and $y$ denote the row and column locations respectively in a map of pixels. A feature image $ \tilde{\mathcal{B}}^{(i)} $, where $1 \leq i \leq \tilde{B}$, defined on $ \mathbb{R}^2$ is a random image which maps each site location to an intensity value $ p \in \mathbb{R}$:
\begin{equation}
\tilde{\mathcal{B}}^{(i)}: \mathbb{R}^2 \to \mathbb{R},  \text{for} \ 1 \leq i \leq \tilde{B} 
\end{equation}
A multi-band image is a $\tilde{B}$-tuple of bands $(\tilde{\mathcal{B}}^{(1)},\tilde{\mathcal{B}}^{(2)},\dotsc,\tilde{\mathcal{B}}^{(\tilde{B})})\in \mathbb{R}^{\tilde{B}}$. Since the multispectral bands and the slope band were collected at different times of the year, in order to gain robustness against the changes in lighting conditions at different times of data collection, the data is transformed via the band difference ratio \citep{Marchisio} as band transformation:

\begin{equation}
\mathcal{D}: \mathbb{R}^2 \to \mathbb{R}^2
\end{equation}
\begin{equation}
\mathcal{D}: (x,y; \tilde{\mathcal{B}}^{(1)},\dotsc,\tilde{\mathcal{B}}^{(\tilde{B})}) \mapsto (x,y; \mathcal{B}^{(1)},\dotsc, \mathcal{B}^{(B)})
\end{equation}
where 
\begin{equation}
\mathcal{B}^{(k)}=\frac{\tilde{\mathcal{B}}^{(i)}-\tilde{\mathcal{B}}^{(j)}}{\tilde{\mathcal{B}}^{(i)}+\tilde{\mathcal{B}}^{(j)}},\text{ for all } i>j, \text{ and } 1 \leq k \leq \binom{\tilde{B}}{2}.
\label{eqn:bdr}
\end{equation}

Our result in Section 4 is based on using $\tilde{B}=9$ and $B=\binom{9}{2} = 36$ band difference ratios for our data analysis. In Section 4.3, we compare the performances using other dimensions of band difference ratios. Note that we can also use the identity function as a band transformation or a combination of band difference ratios and identity function.

\subsection{Annuli Method in Image Processing}\label{sec:imgProc} 

The image processor \citep{Schowengerdt2006} is a function
\begin{equation}
f:\mathbb{R}^2 \to \Re^{\tilde{d}},
\label{eq:imgProcessor} 
\end{equation}

\begin{equation}
f:(x,y;\mathcal{B}^{(1)},\dotsc,\mathcal{B}^{(B)}) \mapsto \mathbf{w}
\end{equation}
that takes a location and $B$ image bands and returns a vector. This function returns local statistics about the image for each location. We suppose that 
\begin{equation}
f(x,y;\mathcal{B}^{(1)},\dotsc,\mathcal{B}^{(B)})=
\begin{pmatrix}
f_1(x,y;\mathcal{B}^{(1)})\\
f_2(x,y;\mathcal{B}^{(2)})\\
\vdots \\
f_B(x,y;\mathcal{B}^{(B)})\\
\end{pmatrix} \in \Re^{\tilde{d}},\quad \forall (x,y),\mathcal{B}^{(1)},\dotsc,\mathcal{B}^{(B)}
\label{eq:imgProcessor2}
\end{equation}
where each component $f_b:\mathbb{R}^2 \to \Re^{\tilde{d}/B}$ depends only on band $b$ for each $b \in \{ 1, \dotsc, B\}$. Further we assume that all the functions $f_b$ are identical, so that the same statistics are computed for each image. 

\begin{figure} 
  \centering
  \includegraphics[width=4in]{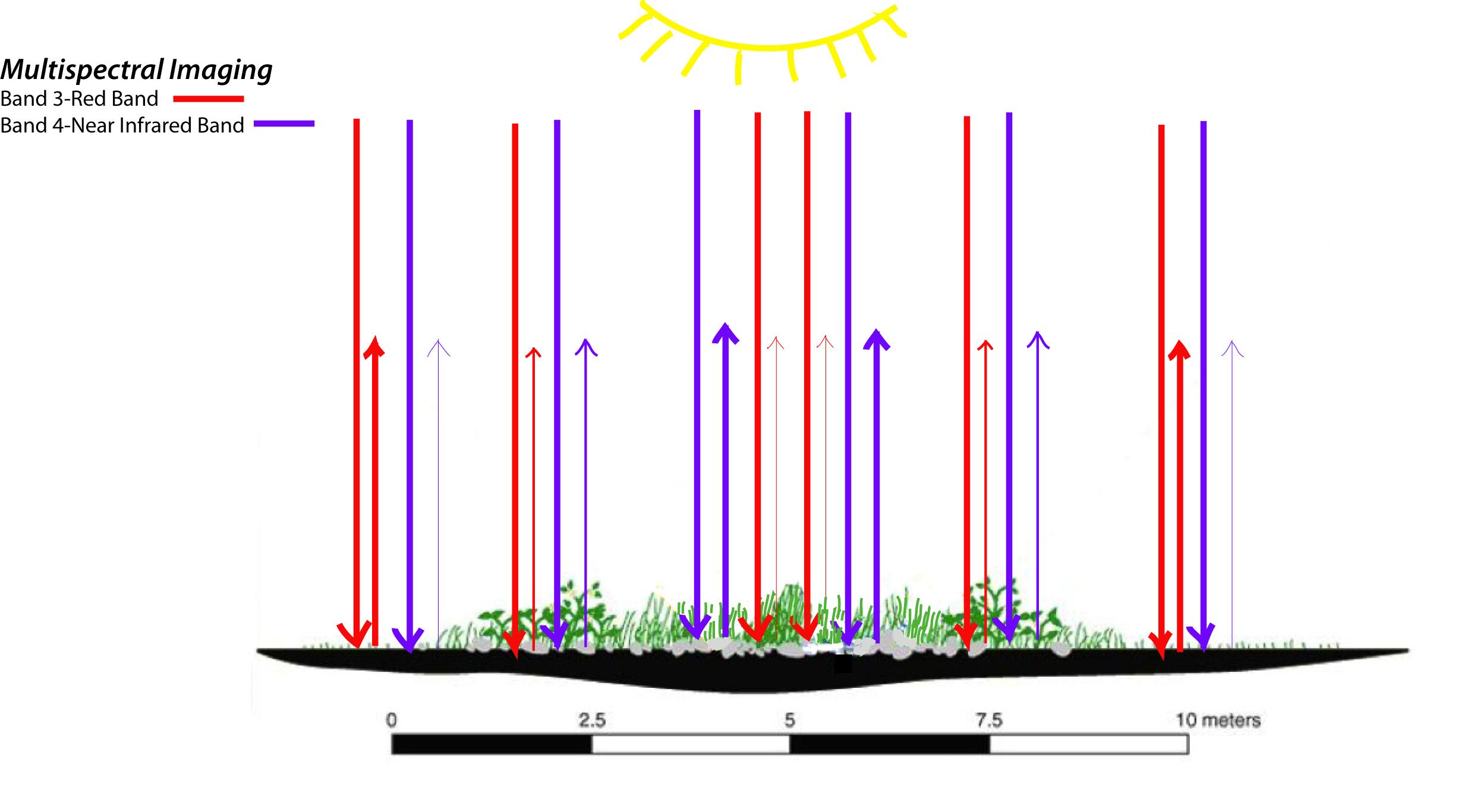}
  \caption[]
   {At typical sites in our test area: There is a core of longer grasses in the middle of the site and further away from the center, vegetation is more sparse. Band 3, red, is almost completely absorbed by healthy, lush vegetation, while Band 4, near-infrared, is strongly reflected by healthy, lush vegetation. The bands reflections thus are different at different parts of the site, but even more different from the surrounding vegetation. This is the reason that we use the annuli approach.}
   \label{fig:reflection}
\end{figure}

In our study, the sensor bands reflect differently around the archaeological sites, seen in Fig [\ref{fig:reflection}]. To capture the differentiating characteristics, we use a particular imaging processing method based on statistics of pixels in annuli centered
at each site and compute two robust statistics.
Let 
\begin{equation}
A_{s}(r^{(in)},r^{(out)})=\{s'\in\mathbb{R}^2: r^{(in)}\leq \|s-s'\|< r^{(out)}\}
\label{eq:annuli}
\end{equation}
be the set of coordinates in the annulus centered at $s\in\mathbb{R}^2$ with inner radius $r^{(in)}$ and outer radius $r^{(out)}$.
We define $\nu:\mathbb{R}^2 \mapsto\Re^{30}$ and $\delta:\mathbb{R}^2 \mapsto\Re^{30}$ by 
\begin{align} \label{eq:medianMAD}
\nu_i(s,\mathcal{B})&=\mathrm{median}\{ \mathcal{B}_{s'}: s'\in A_{s}(r_{i}^{(in)},r_{i}^{(out)})\} \\
\begin{split}
\delta_i(s,\mathcal{B})&=\mathrm{MAD}\{ \mathcal{B}_{s'}: s'\in A_{s}(r_{i}^{(in)},r_{i}^{(out)})\}\\
&=\mathrm{median}\{|\mathcal{B}_{s'}-\nu_i(s)|: s'\in A_{s}(r_{i}^{(in)},r_{i}^{(out)}) \}
\end{split}
\end{align}
for $i\in\{1,2,\dotsc,30\}$ where MAD denotes the sample median absolute deviation.

The inner and outer radii ranges we consider are given in this table:
\begin{center}
\begin{tabular}{|c||rrrrr|rrrrr|rrrrr|}\hline
$i$ & 1 & 2 & 3 & \dots & 10 & 11 & 12 & 13 & \dots & 20 & 21 & 22 & 23 & \dots & 30 \\\hline
 $r^{(in)}$ & 0 & 3 & 6 & \dots & 27 & 0 & 5 & 10 & \dots & 45 & 0 & 7 & 14 & \dots & 63 \\
$r^{(out)}$ & 2 & 5 & 8 & \dots & 29 & 4 & 9 & 14 & \dots & 49 & 6 & 13 & 20 & \dots & 69 \\\hline
\end{tabular}
\end{center}
Since the annuli overlap and cover a relatively large area, the statistics for these ranges will likely
contain some redundant information so the feature extraction step will be useful to reduce this redundancy
while maintaining the components that discriminate between classes. 

\subsection{Feature Extraction and Classification}\label{sec:featExtr_cls}

Each location is now represented by a vector in $R^{\tilde{d}}$ (with $\tilde{d}=60B$). 
The dimension $\tilde{d}$ is much larger than the sample size $n$.
This often leads to ``the curse of dimensionality'' \citep{Duda,tan}, a phenomenon in which the available sample sizes are insufficient to build reasonable models of the high dimensional data. Such high dimensionality can also result in complicated computational issues and reduces classification
accuracy. Feature extraction eliminates irrelevant features and reduces noise in
the data set. Moreover it converts high dimensional feature space
to a lower dimension that captures most of the variation of the original
data. Mathematically, the feature extractor $g$ is a function such that:

\begin{equation}
g: \mathbb{R}^ {\tilde{d}} \to \mathbb{R}^{d} \text{, with } d<\tilde{d}.
\end{equation}

In this paper, principal components analysis (PCA) is used for feature extraction. The classifier $h$ is a function
\begin{equation}
h:\Re^d\to [0,1].
\end{equation}
In our approach, $ g $ and $ h $ are very much intertwined in the sense that, $ h $ is applied in the procedure of choosing PCA dimension $d$, and the optimal $d^{*}$ is applied for the data before classification $ h $ is done. Further interaction between $ g $ and $h$ is through the nested loops in leave-one-out cross validation. An algorithmic explanation of this procedure is presented in Algorithm 1.   

The classifier we choose is linear discriminant analysis (LDA). The predicted class is given by $\mathcal{Y}_{pred}=\mathbb{I}\{h>\tau\}$, i.e, the site has archaeological significance if the estimated posterior probability is greater than the threshold $\tau\in[0,1]$. In Section 5, we discuss using another classifier $(k,l)$-nearest neighbor rule \citep{dev} and compare the PCA errors of both classifiers.

Receiver operating characteristic (ROC) curves are parametrized by the the threshold $\tau$. For each value of $\tau$, we plot the percentage of false positives on the $x$-axis and the percentage of true positives on
the $y$-axis. The area under the ROC curve (AUC) is a summary statistic which averages the true positive over all choices for the false positive rate.

\begin{algorithm}

\caption{Use $APM_{enhanced}$ to classify archaeological sites} 

\begin{algorithmic}[1] 

\State Apply band difference ratio to multispectral band data

\State Apply the annuli technique $f$ and calculate the medians and MADs for each annulus
\For{$s = 1 : n_{test} $}
	\For {$d = 1 : (n_{test}-1) $} 
		\State Apply $h\circ g_{d(s)}$ on the training set
		\EndFor		
		\State Calculate misclassification error $\epsilon_{d(s)}$ 
		\State Pick $d^{*}_{(s)}=argmin_{\{1,\dotsc,d\}}\epsilon_{d(s)}$.

\State Apply $h\circ g_{d^{*}_{(s)}}$ to test $s$ and get the posterior probability 					    
\EndFor

\end{algorithmic}
\end{algorithm}

\section{Results from the Ft.\ Irwin Site Research}\label{results}

\subsection{Assessing the Performance on Training Data set}
We analyze $APM_{enhanced}$ on the multi-band dataset. There are $n_1=37$ archaeological sites in Class $1$. For Class $0$ data, we select $n_0=100$ surveyed locations uniformly random \footnote{Surveyed locations are defined as locations where archaeologists surveyed but discovered nothing. Thus it is safe to assume they are non-sites.} and ensure these regions are far from known sites. Each site used in this analysis was examined by Cultural Site Research and Management (CSRM) under a contract with the Department of the Army to determine the eligibility of these for sites for inclusion in the National Register of Historic Places. This investigation was done under the direction of Douglas C. Comer, during 2011 and 2012. Precise locations were recorded with geographical positioning system (GPS) equipment, and site types were determined, and reviewed, by archaeologists who specialize in the Mojave Desert region.
The APM for Ft.\ Irwin \citep{Ruiz}
 is specifically defined as
\begin{equation}
APM_{conventional}: \mathbb{R}^2 \to \{0, 0.2, 0.4, 0.6, 0.8, 1\}
\end{equation}
where $0$ indicates the site is least likely to be archaeological and $1$ the most. We further consider a convex combination $APM_{\gamma}$. For each value of $\gamma\in[0,1]$, 
\begin{align}
APM_{\gamma}  &= (1-\gamma)APM_{conventional}+\gamma APM_{enhanced}.
\end{align}

\begin{figure} 
  \centering
  \includegraphics[width=4in]{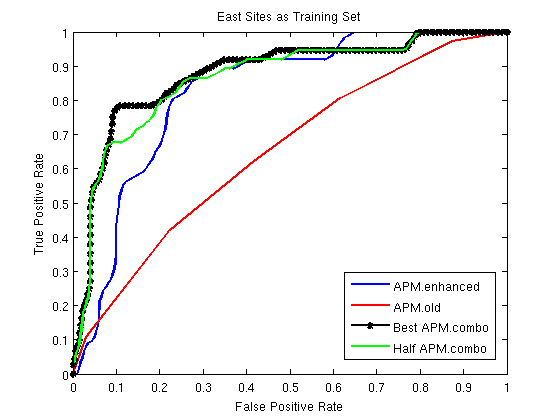}
  \caption[ROC for Training on Eastern Ft.\ Irwin Sites]
   {ROC for Training on Eastern Ft.\ Irwin Sites}
   \label{fig:train_east}
\end{figure} 

\begin{figure} 
  \centering
  \includegraphics[width=4in]{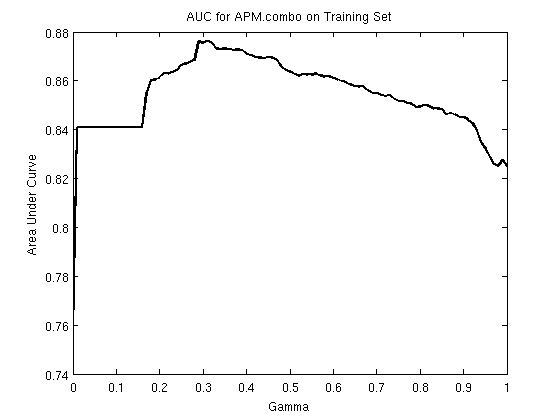}
  \caption[AUC for Training on Eastern Ft.\ Irwin Sites]
   {AUC for Training on Eastern Ft.\ Irwin Sites. Notice that there is a jump at $0$. That is because $APM_{enhanced}$ acts like a tiebreaker and rescales the discretized values from $APM_{conventional}$.}
   \label{fig:auc_train}
\end{figure} 
Figure [\ref{fig:train_east}] shows 
the receiver operating characteristic (ROC) \citep{bradley1997use}
 curve of lithic sites
corresponding to four classifiers: $APM_{conventional}$ ($\gamma=0$), $APM_{enhanced}$ ($\gamma=1$),
$APM_{\gamma=0.5}$, and the optimal combined model $APM_{\gamma=\gamma^*}$. Figure [\ref{fig:auc_train}] shows how we choose the best $\gamma$ to get $APM_{\gamma=\gamma^{*}}$ and $\gamma^{*}=0.3$. For different choices
of false positive percentage (or true positive percentage) we see that $APM_{\gamma}$ typically improves performance over using either the $APM_{conventional}$ and $APM_{enhanced}$ alone.
The consistent superior performance of $APM_{\gamma}$ to $APM_{conventional}$ indicates that including our method increases prediction power over using the $APM_{conventional}$ alone. Also the false negative rate is lower in our enhanced model than in $APM_{conventional}$.

\subsection{ Testing on the West of Ft.\ Irwin Dataset}
We train on the eastern sites using our method and test on the western Ft.\ Irwin sites. There are $ n_{1} =49$ archaeological sites and $ n_{0}=100 $ non-archaeological sites chosen at random from the western region surveyed map. We ensure that these non-sites are selected at least $ 200 $ meters away from the found and not-found sites. $ APM_{enhanced}$ is trained on the eastern region and tested on the western region.
\begin{figure}
  \centering
      \includegraphics[width=0.5\textwidth]{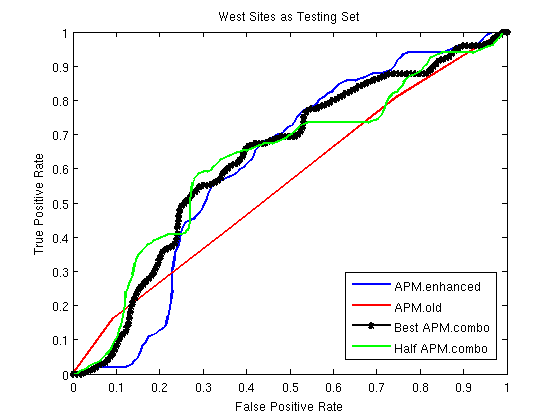}
  \caption{ROC curve for testing on the western lithic sites}
   \label{fig:test_west}
\end{figure}

\begin{figure} 
  \centering
      \includegraphics[width=0.5\textwidth]{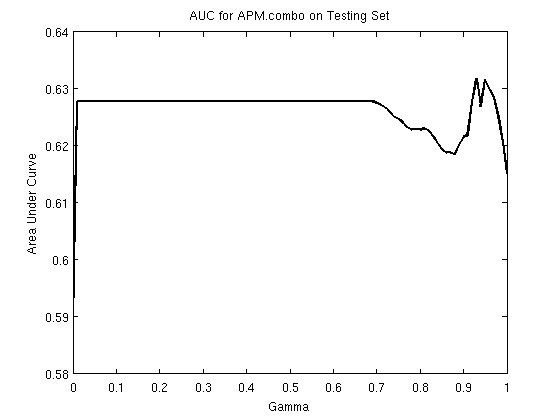}
  \caption{AUC for testing on the western sites. The first jump is a discontinuity because $APM_{enhanced}$ acts like a tiebreaker. The longer flat line in AUC from $\gamma \in [0,0.7]$ (compared to $\gamma \in [0,0.23]$ in Fig [\ref{fig:auc_train}]) is due to the fact that the scores from testing have a smaller range.}
  \label{fig:auc_west}
\end{figure}

Figure [\ref{fig:test_west}] shows that our method in general outperforms APM, except from when false positive rates in $[0, 0.23]$. The convex combination model at the optimal $ \gamma $ beats APM starting from false positive rate at $0.13$, indicating that including our method improves classification power. We also see that our method achieves higher true positive rate at lower false negative rate than the APM from $0.13$ to $1$. Since we are more interested in achieving lower false negative rate, we mainly concerned the false positive rate ranging from $0.15$ to $1$. Indeed, at these rates, our method outperforms the $APM_{conventional}$. Figure [\ref{fig:auc_west}] shows how the AUC of $APM_{\gamma}$ is varied for different choices of $\gamma$. We see that $\gamma$ closer to $1$ has higher AUC, which indicates weighing more $APM_{enhanced}$ increases classification accuracy.

\section{Discussion}\label{Discussion}
\begin{figure}
  \centering
  \includegraphics[width=4in]{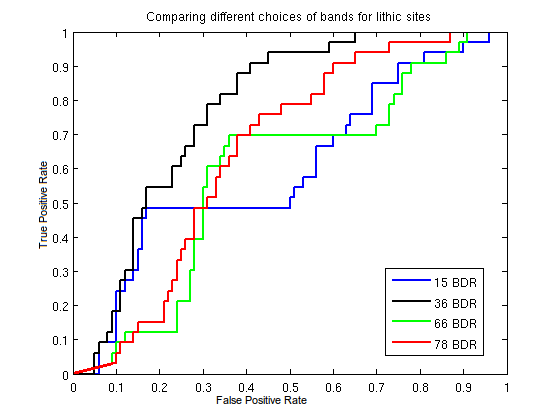}
  \caption[Compare choices of band under the best corrections]
   {Compare dimensions of band difference ratios using $APM_{enhanced}$}
   \label{compare_bdr_dim}
\end{figure}

In Section 3.1, we describe the band transformation function in $APM_{enhanced}$ and, in particular, use 36 band difference ratios. We also examine many sensor bands: eight multispectral bands, four KTTs, slope, one NDVI, where the four KTTs and one NDVI \citep{yarbrough2005using} were not independent from, moreover generated by the eight multispectral bands. We consider cases of 15, 36, 66 and 78 band difference ratios, where the combinations are chosen by their archaeological importance and are demonstrated in Table [2]. We show that using the 36 dimensions produces the highest classification accuracy as in Fig [\ref{compare_bdr_dim}]. Since the imaging technique extends each band to 60 dimensions and with only above 100 training samples, higher dimensional features contain too much noise in the dataset while low dimension features do not capture enough information. Nevertheless, if given more training samples, the analysis using 66 dimensions may generate better results.
\begin{table}
\caption{Compare Band Difference Ratio Dimensions}
\begin{tabular}{ | c | p{7cm} | c | c |} \hline
\textbf{Number ofBDR} & \textbf{Names of bands} & \textbf{Perform} \\\hline
\textbf{15 BDR} & Slope, NIR-1, red edge, brightness, greeness, wetness & Worst \\\hline
\textbf{36 BDR}& Slope, 8 WorldView-2 bands & Best \\\hline
\textbf{66 BDR} & Slope, 8 WorldView-2 bands, brightness, greenness, wetness & Second best \\\hline
\textbf{78 BDR} & Slope, 8 WorldVIew-2 bands, 4 KTT bands & Third best \\\hline
\end{tabular}
\label{table_bdr_dim}
\end{table}

In our classification scheme framework, we compare performances of LDA versus $(k,l)$-nearest neighbor rule \citep{dev} given by:

\begin{equation}
h(n) = \left\{
  \begin{array}{l l}
    1 & \quad \text{if $\sum_{i=1}^n \mathcal{Y}_i \geq l$}\\
    0 & \quad \text{if $\sum_{i=1}^n \mathcal{Y}_i \leq k-l$ } \\
    -1 & \quad \text{otherwise.}

  \end{array} \right.
\end{equation}
We have empirical evidence that the PCA error of $(k,l)$-NN is lower than that of LDA, since the best PCA dimension is much smaller using $(k,l)$-NN. However, by the bias-variance trade-off, the predicting performance of $(k,l)$-NN does not work as well as for LDA. However both classifiers still achieve higher classification accuracy than $APM_{conventional}$.

In our framework, the criteria for selecting PCA dimension $d^{*}$ is based on lowest error rate $\epsilon$. But we can also consider the ratio of the eigenvalues of the covariance matrix over the sum of all the eigenvalues. In that sense, we are choosing $d^{*}$ based on percentage of the variation of data and a common threshold is $95\%$. In this paper, the area under the ROC curve is used as a summarizing number for testing the performance of our algorithm. However, we can also consider using precision-recall as another measure to evaluate the performances.

Our improved archaeological predictive methodology $APM_{enhanced}$, specified by the four components: band transformation to data, image processor, feature extractor, and classifier, is applied to archaeological site discovery using multispectral and topographical imagery data. $APM_{enhanced}$ is trained on the eastern region of Ft.\ Irwin and tested on the western region of Ft.\ Irwin. $APM_{enhanced}$ demonstrated not only lower error rate but also lower false negative rate. With low false negative rate, road constructions, or other activities that disturb the ground including military maneuvers have less chance building through or ruining an archaeological site nearby, thus saving tremendous costs. 

$APM_{conventional}$, a standard means for identifying archaeological sites, inputs features such as slope, vegetation, elevation and so on. These factors may not be preserved well and the input values are measured by archaeologists investigating in the field. $APM_{conventional}$ predicts on low-dimensional data, usually less than $10$. $APM_{enhanced}$ is a statistical learning approach that applies on multispectral band data acquired from advanced remote sensing technologies and predicts on high-dimensional data. 

$APM_{conventional}$ outputs a few discretized values. The ability to adjust false negative rates is limited because the threshold $\tau$ can only make a difference at few values. $APM_{enhanced}$, on the other hand, outputs much more values so that false negative rates can be adjusted lower. At the same threshold $\tau$, $APM_{enhanced}$ has lower false negative rate than $APM_{conventional}$. 
$APM_{conventional}$ returns likelihoods of whether a region contains archaeological sites so sites within a region have the same $APM_{conventional}$-probabilities. $APM_{enhanced}$ returns probabilities of single sites as to enhance the accuracies. $APM_{conventional}$ is a map-based approach because a different map may be influenced by a different set of geological and environmental factors.  Hence it has less transferability and requires archaeologists to survey the field, which is costly and time-consuming. Our machine-learning-based $APM_{enhanced}$ approach has operationally useful transferability as shown in this paper. Although $APM_{enhanced}$ and $APM_{conventional}$ have many differences, they are fundamentally used for finding archaeological sites. Combining both methodologies $APM_{\gamma}$ generates even better results. Our analysis strongly suggests that combining statistical learning methods and human professional knowledge can bring forward fascinating outcomes.

\bibliography{bibref}
\bibliographystyle{plainnat}
\end{document}